# Versatile manipulation of low-refractive-index particles using customized optical building blocks

Minru He,[1] Yansheng Liang,[1,*] Xue Yun,[1] Linquan Guo,[1] Tianyu Zhao,[1] and Ming Lei[1,2,**]

[1]MOE Key Laboratory for Non-Equilibrium Synthesis and Modulation of Condensed Matter, School of Physics, Xi'an Jiaotong University, Xi'an 710049, China

[2]State Key Laboratory of Electrical Insulation and Power Equipment, Xi'an Jiaotong University. Xi'an 710049, China

*yansheng.liang@mail.xjtu.edu.cn

**ming.lei@mail.xjtu.edu.cn

**Abstract:** Low-refractive-index (LRI) particles play significant roles in physics, drug delivery, biomedical science, and other fields. However, they haven't attained sufficient utilization in active manipulation due to the repulsive effect of light. Here, we demonstrate the establishment of optical building blocks (OBBs) to fulfill the demands of versatile manipulation of LRI particles. The OBBs are generated by assembling generalized perfect optical vortices based on the free lens modulation (FLM) method, by which the beams' shape, intensity, and position can be elaborately designed with size independent of topological charge. Using the OBBs with high quality and high efficiency, we realized rotating LRI particles along arbitrary trajectories with controllable speed and parallel manipulation of multiple LRI particles. Importantly, we further achieved the sorting of LRI particles by size with specially structured OBBs. With unprecedented flexibility and quality, OBBs provide tremendous potential in optical trapping, lithography, and biomedicine.

## Introduction

"Optical tweezers (OTs)" [1], introduced by Ashkin in 1986, are important tools that use tightly focused laser beams to exert or measure forces/torques on particles in three dimensions. Since then, numerous advancements in this technology have promoted OTs as powerful tools in biology[2,3], physics [4,5], and other microscale research [6]. With the development of beam shaping techniques, a variety of structured beams have been studied, enabling versatile manipulation of micro/nano objects, such as optical vortices (OVs) [7], perfect optical vortices (POVs) [8], Bessel beams [9], and line patterns [10]. The most used objects for OTs are high-refractive-index (HRI) particles, whose refractive index is greater than the surrounding medium's. Low-refractive-index (LRI) particles, such as vesicles [11], droplets [12], and bubbles [13,14], also play significant roles in physics, biomedicine, and biotechnology. For instance, droplets and vesicles have been applied as self-contained and flexible microreactors in drug delivery [15,16] and used to fabricate biopolymer structures in artificial cell synthesis [17–19]. Various LRI subcellular organelles have been observed recently in cell functions [20] and structures [21]. Given the significance of LRI particles, versatile manipulation of such particles will help to investigate their properties, thus providing a better understanding of the particles' interactions in cell biology, biomedicine, biopharmaceuticals, etc.

Unlike HRI particles, LRI particles tend to be repelled away from the bright spot [22], making stable trapping and controllable manipulation of LRI particles a big challenge. The straight strategy to confine LRI particles is creating a dark region with low intensity surrounded by a high-intensity barrier [22–25]. Yet more precise, controllable, and versatile manipulation of LRI particles remains an ongoing challenge. POVs provide a viable solution for LRI particle manipulation, because of ring-like intensity profiles independent of topological charge (TC) [25]. However, in the particle manipulation, the rotation of LRI particles using POVs was limited by the fixed ring trajectory. Inspired by the recently demonstrated "double ring (DR) POVs" with dark ring profiles [26] and GPOVs with arbitrary trajectories [27], double orbit GPOVs with arbitrary profiles will facilitate versatile manipulation of LRI particles.

In this paper, we present constructing optical building blocks (OBBs) to manipulate LRI particles multifunctionally. The OBBs are created by assembling multiple structured beams generated with the free lens modulation (FLM) method [28]. Using this method, we produced various types of OBBs for versatile manipulation of LRI particles, including double orbit GPOVs, GPOVs array, and other specially structured patterns for size sorting. We readily achieved velocity-controlled rotation along arbitrary trajectories, parallel dynamic manipulation, and sorting of LRI particles using the produced OBBs. To our knowledge, this is the first time that the versatile manipulation of LRI particles has been demonstrated. The OBBs provide high-flexibility and high-quality of beam profiles. Therefore, we expect the OBBs to significantly expand the application potential of LRI particles, facilitating their enhanced contributions to biology, medicine, physics, and other fields.

## Methods

### Light fields generation

The idea of generating OBBs is based on the FLM method, which functions as an "optical pen" [28] to shape arbitrary beam patterns with controllable OAM. The transmission function of the digital lens designed by the FLM method can be written as

$$t_l^{\mathrm{G}} = P(r,\varphi) e^{-ik(r-\rho_0(\varphi))^2/2f(\varphi)} e^{il\varphi} \tag{1}$$

where $(r, \varphi)$ are the polar coordinates, $P(r, \varphi)$ represents the aperture function, $k = 2\pi/\lambda$ refers to the illumination wave number, and $\rho_0(\varphi)$ and $f(\phi)$ represent the shape and focal length of the designed digital lens, respectively. The diffracted light field at the focal plane of the digital lens, according to the Fresnel diffraction theorem, satisfies

$$E(\mathbf{u}_0(\varphi)) = A \frac{e^{ikz}}{i\lambda\varphi} e^{i\pi\lambda^2/\lambda z} \mathbf{F}\left[ e^{ik(r-\rho_0(\varphi))^2/2f\varphi} e^{i\pi r^2/\lambda z} e^{il\varphi} \right] \tag{2}$$

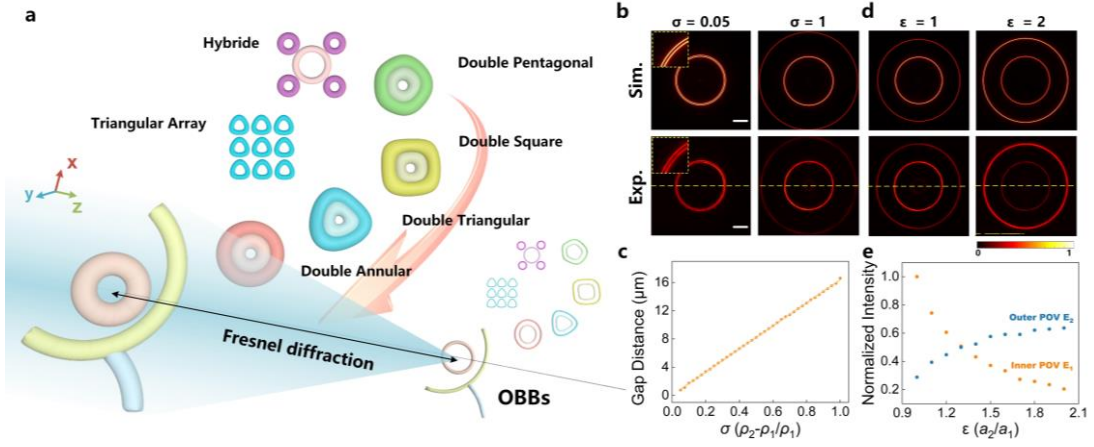

**Fig. 1 Generation of optical building blocks (OBBs). a** Schematic illustration for generating OBBs using the FLM method. OBBs' profiles correspond to free lens models with different colors. **b** Simulated (first row, Scale bar = 10 mm) and experimental (second row, Scale bar: 10 μm) DR POVs, with $\sigma$ = 0.05 and 1. **c** Gap width against $\sigma$ plot, with $\sigma$ from 0.05 to 1. **d** Simulated (first row) and experimental (second row) results, with $\varepsilon$ = 1 and 2. **e** Normalized intensity plot of inner and outer POVs, with intensity ratio $\varepsilon$ changing from 0.05 to 1.

where $\mathbf{u_0} = (\rho_0(\varphi), \varphi, f(\varphi))$ denotes the 3D curve parameterized by the azimuth angle $\varphi$, $\mathbf{F}$[...] represents the Fourier transform. Our previous study [28] has demonstrated the FLM method's capability to generate high-perfection and high-efficiency GPOVs. Furthermore, we assembled multiple GPOVs to generate multifunctional complex light fields and named them "optical building blocks (OBBs)". The assembled light field is written as

$$E(\mathbf{u}(\varphi)) = a_1 E(\mathbf{u}_1(\varphi)) + a_2 E(\mathbf{u}_2(\varphi)) + \cdots + a_n E(\mathbf{u}_n(\varphi)) \tag{3}$$

where $a_1$, $a_2$...$a_n$ represent the amplitude weights of different GPOVs, $E(\mathbf{u}_1(\varphi))$, $E(\mathbf{u}_2(\varphi))$ ... $E(\mathbf{u}_n(\varphi))$ represent the freely modulated light field components. Figure 1a shows the schematic illustration of generating OBBs using the FLM method. The combined multiple free lenses generate the corresponding light fields at the focal plane.

As a typical type of OBBs, the double-orbit OBBs (DO OBBs) are generated by combining two GPOVs. By doing so, the DO OBBs have a dark intensity gap between the two GPOVs, by which the LRI particle can be confined in the gap. Driven by the transverse force arising from the phase gradient, the particle will be transported along the beam trajectory. To investigate the trapping and manipulation of LRI particles using DO OBBs, we take DR POVs as examples by combining two ring profile GPOVs. The DR POVs have tunable gap widths between the two rings, exhibiting greater flexibility than the previously reported ones [26]. We theoretically and experimentally investigated DR POVs by changing the gap width between the two rings and the intensity distribution (Figs. 1b-e). Our experimental setup comprises a linearly polarized near-infrared (NIR) laser (wavelength = 1064 nm, Connet Laser Technology, Shanghai, China), a pure phase spatial light modulator (SLM, Pluto-NIR-II, HOLOEYE Photonics, Berlin), and a high numerical aperture (NA) oil-immersion objective (100×/NA=1.4, Nikon, Tokyo) (See Supplementary Section 1 for details). To be consistent with the experiment, for simulation we set the light wavelength to 1064 nm, the TC to 10, the input Gaussian waist to $w_0$ = 2.6 mm, and the focal length of the digital lens to $f_0 = 200$ mm. The size of the simulated DR POVs in the focal plane of the digital lens is 100 times that in the focal plane of the objective (100×). The gap width $d_{gap}$ of the two rings can be controlled by changing the radius ratio $\sigma$, defined as

$$\sigma = (\rho_2 - \rho_1)/\rho_1 = d_{gap}/\rho_1 . \tag{4}$$

where $\rho_1$ and $\rho_2$ denote the radius of the inner and outer rings. Setting $\rho_1 = 1.7$ mm, the simulated and experimental gap widths are 81.2 and 0.75 μm for $\sigma = 0.05$, and 1.67mm and 16.15 μm for $\sigma = 1$, respectively (Fig. 1b). As shown in Fig. 1c, the experimental gap width changes linearly with the ratio $\sigma$. Notice that while changing the gap width of the two rings, the intensity difference gets pronounced, which will influence the uniformity of the light fields and the trapping performance. This problem can be overcome by changing the amplitude ratio of the two rings, i.e.,

$$\varepsilon = a_1/a_2 . \tag{5}$$

We performed a comparative analysis by varying $\varepsilon$ from 1 to 2 while setting $\sigma = 1$ (Figs. 1d and 1e). Results show that the maximal intensity of the outer ring can be changed from 28% to 315% of the maximal intensity of the inner one by setting $\varepsilon$ from 1 to 2 without degrading the beam quality. Therefore, the spacing and intensity of the two rings can be freely controlled using the FLM method, which provides remarkable flexibility and controllability in manipulating LRI particles.

## Dynamic analysis

The reported OBBs are expected to be powerful tools for manipulating LRI particles. To validate the trapping performance of such beams, we first numerically calculated the transverse force acting on the particle by DR POVs. We selected hollow glass microspheres with a filling fraction of 91% air as LRI candidates for trapping. For simplicity, we calculated the effective refractive index $n_p$ of hollow glass microspheres and got $n_p = 1.1$ [29] (see Supplementary Section 2 for details). In the simulation, we set the particle's radius $r_p$ to 3 μm, the refractive index of the surrounding media $n_0$ to 1.33, and the focusing NA to 1.4. Figures 2a and 2b show the vector maps of the force experienced by an LRI particle in DR POVs ($\rho_1 = 15.3$ μm, $\rho_2 = 10.7$ μm, $\varepsilon = 1.46$) with TC= 25 and -25, respectively (see Supplementary Section 2 for details). In Fig. 2b, the radial force plot along the dotted line with TC = -25 shows the particle is confined at the positions E1 and E2 between the two rings with F=0 and negative slopes. At these equilibrium positions, the particle will encounter azimuthal force and rotate in the gap between the two rings due to the transfer of OAM. First, we investigated the impact of the gap width on the rotation performance. As shown in Fig. 2c, the LRI particles with $r_p = 2.5, 3$, and 3.5 μm reach the maximal azimuthal force of 2.6, 3.4, and 4.2 pN when the gap width is set to 3.44, 4.20, and 4.96 μm the best width, respectively. Deviating from these values, the force will decrease significantly. Therefore, the gap width should be appropriately set to match the particle size for efficient rotation, which would be achieved assuming that the azimuthal force is larger than 90% of the maximal force. For $r_p = 2.5, 3$, and 3.5 μm, the proper range of the gap width for efficient rotation is [2.96 – 3.84], [3.68 – 4.76], and [4.44 – 5.52] μm, respectively (Table 1). Figure 2d plots the azimuthal force against the TC ranging from 10 to 50 for LRI particles with $r_p$ = 2.5, 3, and 3.5 μm at the equilibrium positions when setting the ideal gap width for each particle. From Fig. 2d, we find that the azimuthal force varies linearly with the TC, regardless of the particle size.

**Table 1. Simulated ideal gap widths of DR POVs and azimuthal forces applied on LRI particles.**

| Particle radius (μm) | Ideal gap width (μm) | Maximal azimuthal force (pN) | Suitable gap width range (μm) |
|---|---|---|---|
| 2.5 | 3.44 | 2.59 | 2.96 – 3.84 |
| 3 | 4.20 | 3.41 | 3.68 – 4.76 |
| 3.5 | 4.96 | 4.17 | 4.44 – 5.52 |

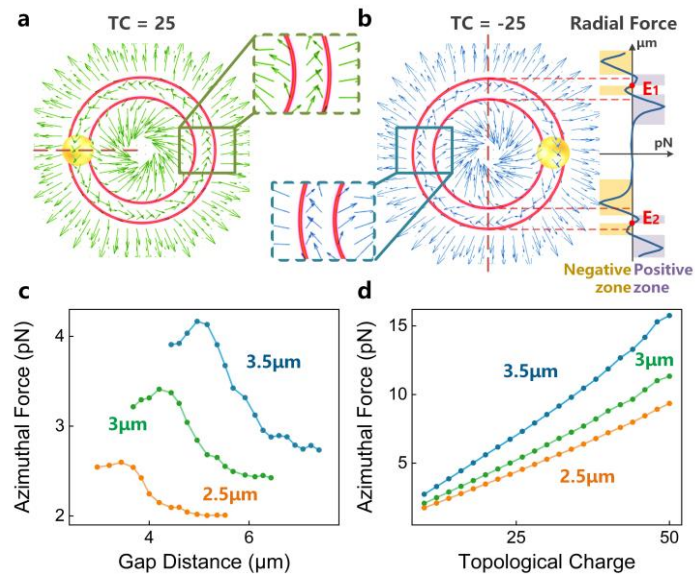

**Fig. 2 Simulated dynamic analysis of LRI particles in DR POVs. a** Vector map of simulated transverse force experienced by hollow glass sphere with TC = 25. **b** Vector map of simulated transverse force and distribution of radial force experienced by hollow glass sphere with TC = -25 in the focal plane. **c** Azimuthal force against the gap width on the LRI particles with $r_p$ = 2.5, 3, and 3.5 μm at the equilibrium position with TC = 15. **d** Azimuthal forces with ideal gap width against the TC on the LRI particles with $r_p$ = 2.5, 3, and 3.5 μm at the equilibrium position.

# Results

We achieved versatile manipulation of LRI particles, including quantitative rotation along arbitrary trajectories, multi-particle dynamic manipulation, and sorting by size utilizing diverse structured OBBs.

## Double orbit rotation

Liang et al. first utilized POVs to manipulate LRI particles by introducing "an extra point trap at the center of the vortex" [25] to improve the rotation performance. However, the trajectories are limited to the ring profiles dependent on the LRI particle size. In contrast, the DO OBBs with tunable gap width and ring size can drive particles along adjustable trajectories.

In our experiments, hollow glass microspheres (Im30k, 3M, America) with size ranging from 5 to 40 μm were selected as the trapping LRI candidates. Figure 3a provides screenshots and time-lapse images of the captured LRI particles with radius $r_p$ = 5.3 μm in DR POVs with TC = 28, the gap width $d_{gap}$ = 7 μm. We first investigated the linear rotation of the particles with radius $r_p$ = 3.1, 4.7, and 5.3 μm (see Visualization 1, Fig. 3b and Table 2) driven by DR POVs consisting of a fixed outer ring with $\rho_1$ = 42.3 μm and an inner ring with tunable radius. Besides, the weights are changed accordingly to ensure the two rings have the same intensity maxima (Eq. 7). We can first conclude from the results shown in Fig. 3c that the gap width significantly impacts the rotation efficiency. Changing the gap width, each particle reached a maximum rotation rate. Specifically, the particles with $r_p$ = 3.1, 4.7, and 5.3 μm got the maximal rotation rate of 0.30, 0.42, and 0.50 Hz when $d_{gap}$ is 4.5, 6.0, and 7.1 μm, respectively. Note that the derailment occurred for particles with $r_p$ = 3.1, 4.7, and 5.3 μm when $d_{gap}$ was reduced to 3.1, 5.2, and 6.0 μm, respectively (Table 2). This phenomenon can be explained by the repulsive effect of the light on the LRI particle. The experimental results shown in Fig. 3c and Table 2 are consistent with the theoretical predictions shown in Fig. 2c. Repeated experiments lead to a rough conclusion that the optimal rotation performance will be realized with $d_{gap}/r$ set in the range of [1.2, 1.4] and particles tend to derail with $d_{gap}/r$ < 1.1 (Figs. 2c and 3c).

Table 2. Experimental ideal gap widths of DR POVs and corresponding rotation rate of LRI particles.

| Particle radius (μm) | Ideal gap width (μm) | Rotation rate (Hz) | Derailment gap width (μm) |
|---|---|---|---|
| 3.1 | 4.5 | 0.3 | 3.1 |
| 4.7 | 6.0 | 0.42 | 5.2 |
| 5.3 | 7.1 | 0.50 | 6.0 |

Setting the gap width to 7 μm for the particle with radius = 5.3 μm, we then explored the particle dynamics by only changing the topological charge (see Visualization 2). Figure 3d presents the rotation rate against the TC. The linear fitting shows that the rotation rate grows linearly with TC changing from 20 to 42. The rotation speed decreases significantly for TC < 20. In this case, the azimuthal force arising from the phase gradient is so small that the disturbance of the surrounding medium and the friction would have a significant influence on the rotation. For large TCs (i.e., TC ≥ 42) the rotation speed changes slightly because of the degradation of the generated light fields caused the pixelated structure of the SLM for large TC [28].

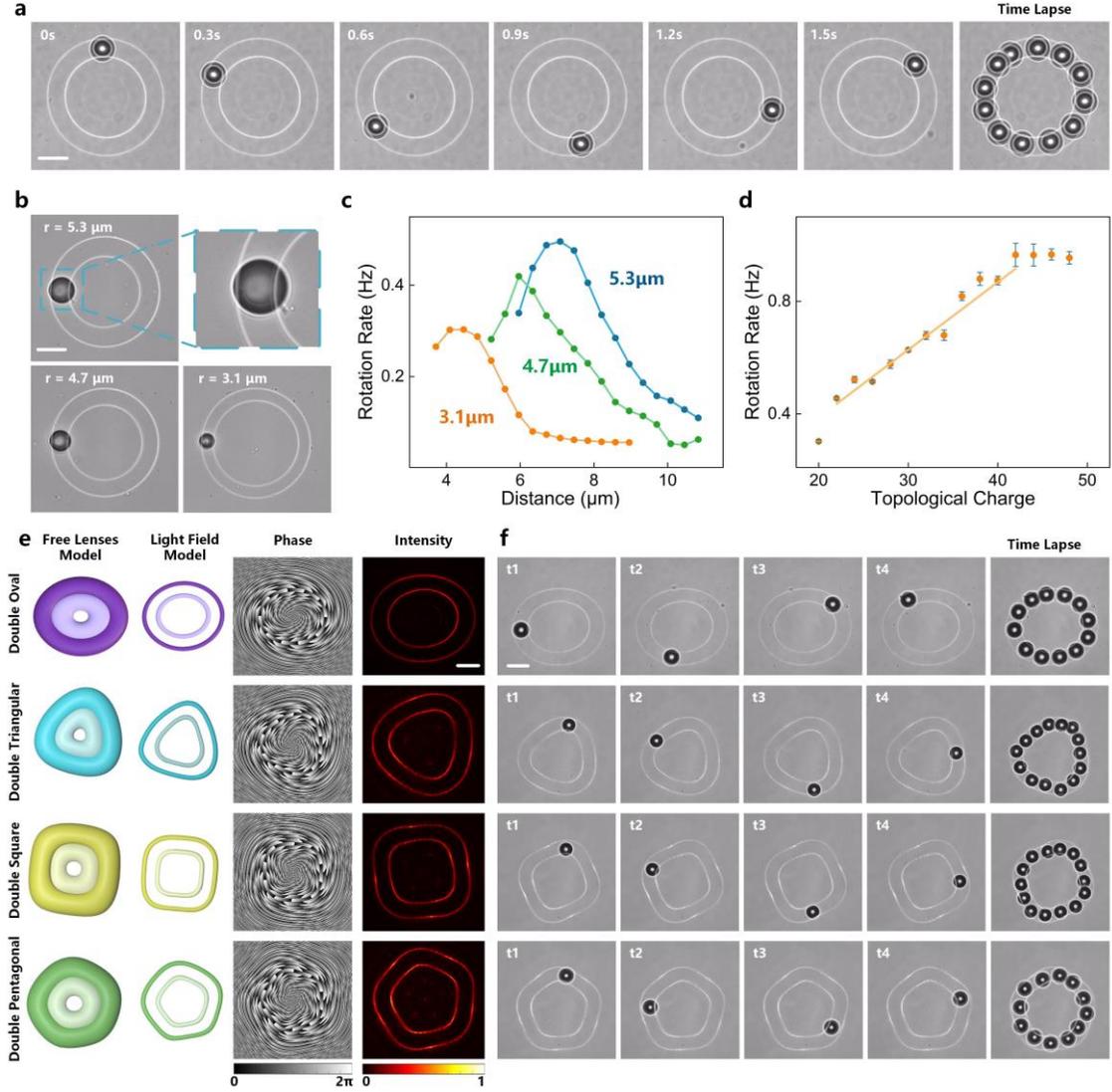

**Fig. 3 Rotation performance of DO OBBs. a** Screenshots (left) and time-lapse image (right) of an LRI particle with $r_p$ = 5.3 µm in DR POVs with TC = 28, $d_{gap}$ = 7 µm. Scale bar = 10 µm. **b** Screenshots and **c** rotation rate against the gap width of LRI particles with $r_p$ = 5.3, 4.7, and 3.1 µm in DR POV with TC = 28 (see Visualization 1). Scale bar = 10 µm. **d** Rotation rate against the TC of the LRI particle with $r_p$ = 5.3 µm in DR POV with $d_{gap}$ = 7 µm (see Visualization 2). **e** Free lens models, simulated light field models, phase maps, and experimentally generated light fields with TC = 10 for double oval, double triangular, double square, and double pentagonal OBBs, respectively. Scale bar: 10 µm. **f** Screenshots (left) and time-lapse image (right) of an LRI particle with $r_p$ = 4.8 µm in DO OBBs corresponding to **e** with TC = 24, $r_p$ = 4 µm (see Visualization 3). Scale bar: 10 µm.

Arbitrary DO OBBs can be created by the elaborate design of the digital lens, which could go far beyond manipulating LRI particles along circular trajectories. For example, we designed polygonal trajectories for the DO OBBs, which satisfy the expression

$$\rho_n(\varphi) = 1 - \frac{1}{p}\cos(q\varphi), \tag{6}$$

where $p$ controls the smoothness of the polygon, and $q$ controls the shape of the curve. Based on Eq.11, we produced four types of DO OBBs: double oval ($p = 5$, $q = 2$), double triangular ($p = 10$, $q = 3$), double square ($p = 15$, $q = 4$), and double pentagonal ($p = 20$, $q = 5$) OBBs. The free lens models, light field models, phase maps (TC = 10), and intensity patterns of the four types of DO OBBs are shown in Fig. 3e. To obtain uniform intensities between the inner and outer rings, we set $\varepsilon = 1.05$. The DO OBBs confined the LRI particles in the gap between the two profiles at the equilibrium positions, and drove them along the shaped orbits with the force along the trajectory (see Visualization 3). Figure 3f provides screenshots of the captured particles and time-lapse images of the manipulation results with TC = 28 and $d_{gap}$ = 5.9 μm, showing that the particle undergoes the double oval, double triangular, double square, and double pentagonal trajectories for the four types of traps with average rotation rates of 0.16 Hz, 0.18 Hz, 0.20 Hz, and 0.17 Hz (Table 3), respectively.

Table 3. The rotation rate of LRI particles in DO OBBs.

| DO OBBs | Parameters | Rotation rate (Hz) |
| --- | --- | --- |
| Double oval | $p = 5$, $q = 2$ | 0.16 |
| Double triangular | $p = 10$, $q = 3$ | 0.18 |
| Double square | $p = 15$, $q = 4$ | 0.20 |
| Double pentagonal | $p = 20$, $q = 5$ | 0.17 |

## Sorting by size

The outstanding flexibility and quality of OBBs provide more possibilities for manipulating LRI particles. By the elaborate design of the patterns of OBBs, we achieved sorting hollow glass spheres by size [30]. The OBBs for sorting are shown in Fig. 4a, which consists of three components: a central POV (structure beam 1, $s_1$: $\rho$ = 9.4 μm, TC = 20) and two truncated off-center circular structural beams (structure beam 2, $s_2$: $\rho$ = 37.6 μm, TC = 20; structure beam 3, $s_3$: $\rho$ = 44.7 μm, TC = -20). The methods for generating off-center and truncated beams are introduced in Supplementary Section 3. With the repulsive effect of light, the narrow channel consisting of $s_1$ and $s_2$ allows only small particles to enter region $z_2$ and prevents large particles from entering. To ensure large particles get across $s_2$ and enter region $z_3$, we set the amplitude weights $a_1$, $a_2$, and $a_3$ of the three components (Eq. 3) to 1.2, 1, and 0.9, respectively.

We designed a simple microfluidic chip for sorting LRI particles (Fig. 4b, see Supplementary Section 4 for details) based on a microscope slide with a flow rate $\geq$ 20 μm s$^{-1}$. The optical sorting of the hollow glass spheres by size with different gap widths is shown in Figs. 4c-e (see Visualization 4). Initially, all the particles will be confined at zone 1, $z_1$, ($t_1$) due to the repulsive force. Setting the smallest gap width of the channel between $s_1$ and $s_2$ $d_{gap}$ to 8.9 μm, the small-sized particles with $r_p$ = 5.4 and 4.4 μm rotate along zone 2, $z_2$, ($t_2$ and $t_3$), while the larger-sized particle ($r_p$ = 6.5 μm) gets across $s_2$ due to the more vital repulsive force of $s_1$ and enter zone 3, $z_3$, ($t_5$ and $t_6$) as shown in Fig. 4c. Changing the gap width, we can realize the sorting of LRI particles by size with various size range. For example, setting $d_{gap}$ = 6.6 μm, the particle with radius of 3.2 and 3.9 μm rotates along the trajectory of $s_2$ and enters region $z_2$, and particles with radii of 4.0 and 6.1 μm enter region $z_3$ by the action of the microflow (Fig. 4d). Then, setting $d_{gap}$ = 4.2 μm, we transported the particle with $r_p$ = 2.1 μm rotates into $z_2$, and deliver the particles with $r_p$ = 2.5 μm $z_3$ (Fig. 4e), as shown in Table 4. To our knowledge, this is the first time optofluidic technology has been applied to sort LRI particles. Since OBBs offer remarkable flexibility and modulation efficiency, we anticipate more complicated optical fields for precise size sorting of LRI particles.

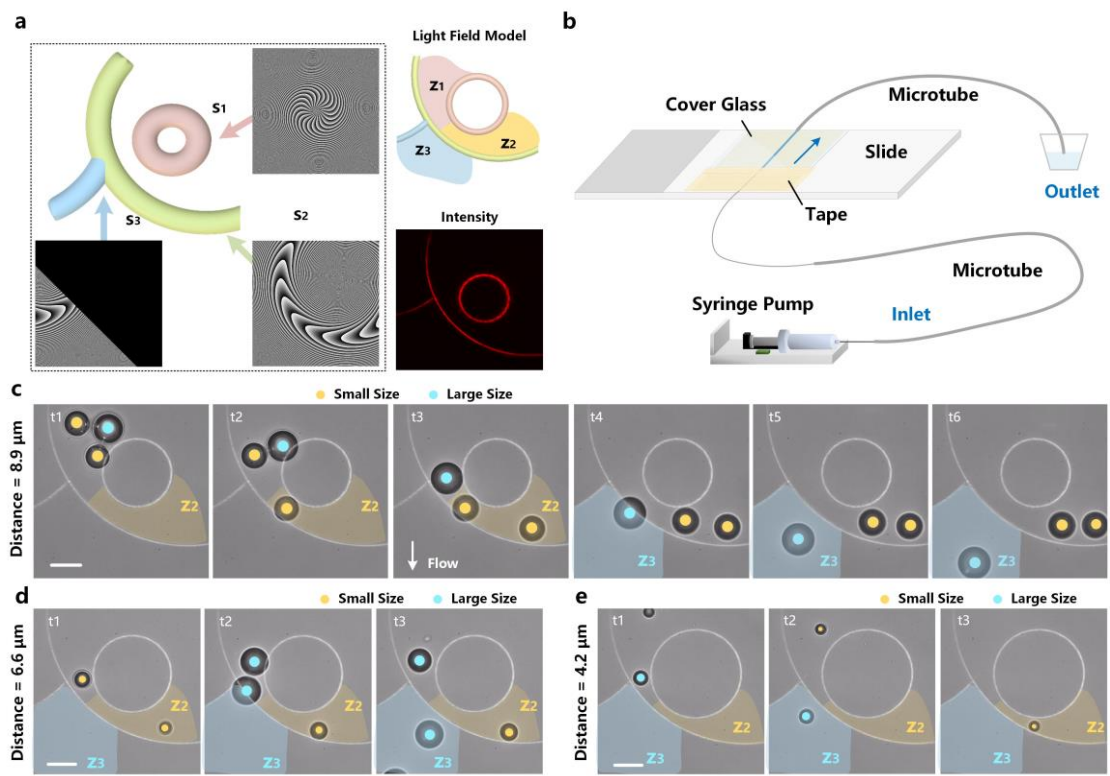

**Fig. 4 Generation of sorting OBBs and sorting experiments by size of LRI particles. a** Free lens models, light field models, and intensity profiles of the OBBs for sorting. **b** Schematic structure for the homemade microfluidic chip (see Supplementary Section 4 for details). Screenshot of sorting of LRI particles by size with setting the gap width $d_{gap}$ to **c** 8.9 μm, **d** 6.6 μm, and **e** 4.2 μm (see Visualization 4). Scale bar: 10 μm.

Table 4. Radii of particles entering regions s2 and s3 at different gap widths

| Gap width (μm) | Radius of LRI particles entering $s_2$ (μm) | Radius of LRI particles entering $s_3$ (μm) |
| --- | --- | --- |
| 8.9 | 4.4, 5.4 | 6.5 |
| 6.6 | 3.2, 3.9 | 4.0, 6.1 |
| 4.2 | 2.1 | 2.5 |

# OBBs Array

In addition to particle rotation, hollow light fields can be used to trap and move LRI particles precisely. In this paper, we generated a series of OBBs arrays including 4 and 9 POVs, intensity-modulated POVs, 9 GPOVs, and hybrid structure beams for Parallel trapping and manipulating of multiple LRI particles (Fig. 5a, see Supplementary Section 3 for off-center OBBs generation).

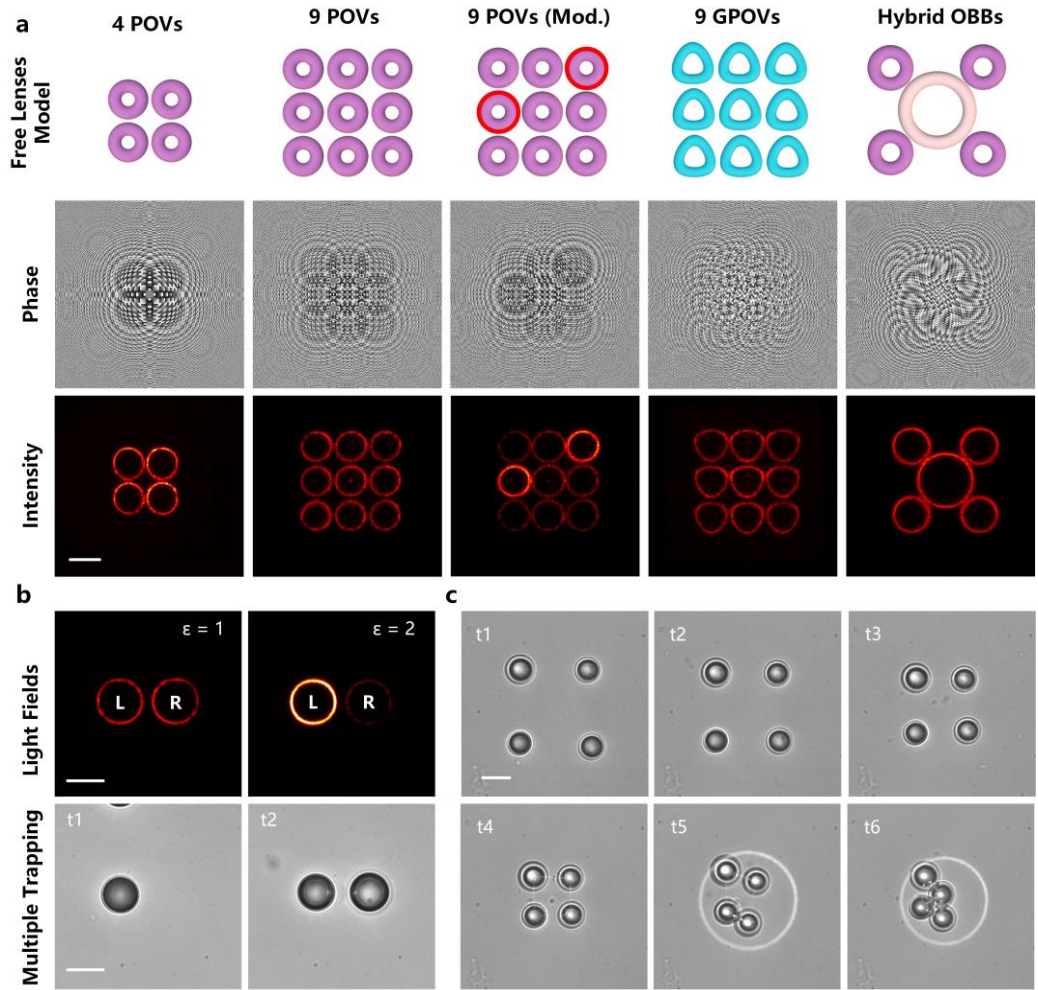

**Fig. 5 Generation and trapping performance of OBBs array. a** Generation of OBBs array. Free lens models, phase map, and intensity profile of 2 POVs, 9 POVs, intensity modified 9 POVs, 9 GPOVs, and hybrid OBBs. Scale bar: 10 μm. **b** Multiparticle trapping procedure with adjustable intensity array. Scale bar: 10 μm. **c** Array trapping and aggregation process of LRI particles (see Visualization 5). Scale bar: 10 μm.

The main challenge in multiple trapping of LRI particles is letting the particles fall into the dark center of all traps through the boundary of the light field while keeping every particle confined by the trap during the capturing process. We demonstrated the adjustable intensity array to solve this problem (Fig. 5b). Taking the 1×2 array as an example, we first set the left ring with much higher intensity than the right one, and trapped an LRI particle with the left ring. Then, we reduced the intensity of the left ring and increased the intensity of the right one, and confined an LRI particle using the right ring while keeping the left ring stably confining the first particle. With this method, we simultaneously captured four LRI particles with $r_p = 3 - 3.5$ μm and moved the four particles closer to each other by decreasing the distance of the trap array (see Visualization 5). Due to the boundaries of the light field, the hollow glass spheres would not directly contact each other as the distance decreased. Therefore, we switched the OBBs array to a single POV with $\rho = 11.48$ μm to let the four LRI particles aggregate (Fig. 5c). Using OBBs, we have successfully achieved the dynamic manipulation and convergence of LRI particles. We anticipate this method will open avenues for future investigations of vesicles and other LRI particles, providing a powerful tool in artificial cell networks.

## Conclusion

In conclusion, the flexibility of the versatile manipulation of LRI particles is derived from OBBs, which are high-quality and high-efficiency assembled structured beams with adjustable TCs, shapes, positions, and intensities. Using the light fields with complex dark regions surrounded by high-intensity barriers, the LRI particles can be confined/driven in arbitrary trajectories. In this paper, we reported the arbitrarily shaped rotation, multiparticle dynamic manipulation, and sorting by size of LRI particles.

The proposed manipulation platform with OBBs can be seen as a functional module. When synergistically coworking with diverse technologies, such as microfluidic techniques and vesicle fusion, the module offers a flexible and efficient approach to realizing more advanced functionalities. For example, the versatile manipulation (such as active selecting, transporting, mixing, and sorting) of the artificial cell components (such as droplets and vesicles, most are LRI particles) will promote the synthesis efficiency and flexibility in biofabrication technology [17]. Incorporating the OBBs into fabrication procedures avoids the high cost and complex microfluidic chips [31]. Also, the OBBs will enable precise control of cell-sized giant vesicles, facilitating complex functionality such as assembling vesicle networks, vesicle fusion, and vesicle communication [32]. Furthermore, the OBBs are expected to be applied to drug delivery [33,34] and microscopy combined with super-resolution techniques [35]. We firmly believe that in the future, the reported OBBs will promote significant advancement in biotechnology, biomedicine, and other fields.

**Funding.** This work was supported in part by the National Key Research and Development Program of China (2022YFF0712500); Natural Science Foundation of China (NSFC) (Nos. 62135003, 61905189, 62205267); The Innovation Capability Support Program of Shaanxi (No. 2021TD-57); and Natural Science Basic Research Program of Shaanxi (Nos. 2020JQ-072, 2022JZ-34).

**Authors' contributions** M.R.H. and Y.S.L. constructed the experimental setup and wrote the code. M.R.H. performed the data analysis. M.R.H. and Y.S.L. wrote the paper. Y.S.L. and M.L. supervised the research project. All authors participated in reading and editing of the final paper.

**Disclosures.** The authors declare no conflicts of interest.

**Data Availability.** Data underlying the results presented in this paper are not publicly available at this time but may be obtained from the authors upon reasonable request.